\documentclass[runningheads]{llncs}

\usepackage[T1]{fontenc}

\usepackage{graphicx}
\usepackage{hyperref}
\usepackage{amsmath}
\usepackage{amssymb}
\usepackage{setspace}
\usepackage{xspace}
\usepackage[capitalise]{cleveref}
\usepackage{listings}
\usepackage{booktabs}
\usepackage{varwidth}
\usepackage{algorithm} 
\usepackage{algpseudocode}
\usepackage{breqn}
\usepackage{mathtools}
\usepackage{placeins}
\usepackage{tabulary}
\usepackage{tabularx}
\usepackage{ltablex}

\usepackage[scaled=.85]{courierten}

\usepackage{caption}
\usepackage{subcaption}

\lstdefinelanguage{json}{
    basicstyle=\normalfont\ttfamily,
    commentstyle=\color{eclipseStrings}, 
    stringstyle=\color{eclipseKeywords}, 
    numbers=left,
    numberstyle=\scriptsize,
    stepnumber=1,
    numbersep=8pt,
    showstringspaces=false,
    breaklines=true,
    frame=lines, 
    string=[s]{"}{"},
    comment=[l]{:\ "},
    morecomment=[l]{:"},
    literate=
        *{0}{{{\color{numb}0}}}{1}
         {1}{{{\color{numb}1}}}{1}
         {2}{{{\color{numb}2}}}{1}
         {3}{{{\color{numb}3}}}{1}
         {4}{{{\color{numb}4}}}{1}
         {5}{{{\color{numb}5}}}{1}
         {6}{{{\color{numb}6}}}{1}
         {7}{{{\color{numb}7}}}{1}
         {8}{{{\color{numb}8}}}{1}
         {9}{{{\color{numb}9}}}{1}
}

\begin{document}
\title{Leveraging Generative AI for Extracting Process Models from Multimodal Documents}

\titlerunning{Extracting Process Models from Multimodal Documents}
%
\author{Marvin Voelter\inst{1,2} \and Raheleh Hadian\inst{1} \and Timotheus Kampik\inst{1} \and Marius Breitmayer\inst{2} \and Manfred Reichert\inst{2}}
\authorrunning{Voelter et al.}
%
\institute{SAP, Berlin, Germany \\
\email{\{marvin.voelter,raheleh.hadian,timotheus.kampik\}@sap.com} \and
Ulm University, Germany \\
\email{\{marius.breitmayer,manfred.reichert\}@uni-ulm.de}
}

\maketitle

\begin{abstract}
This paper presents an investigation of the capabilities of Generative Pre-trained Transformers (GPTs) to auto-generate graphical process models from multi-modal (i.e., text- and image-based) inputs. More precisely, we first introduce a small dataset as well as a set of evaluation metrics that allow for a ground truth-based evaluation of multi-modal process model generation capabilities.
We then conduct an initial evaluation of commercial GPT capabilities using zero-, one-, and few-shot prompting strategies.
Our results indicate that GPTs can be useful tools for semi-automated process modeling based on multi-modal inputs.
More importantly, the dataset and evaluation metrics as well as the open-source evaluation code provide a structured framework for continued systematic evaluations moving forward.
\end{abstract}

\section{Introduction}
\label{sec:introduction}
With the advent of generative artificial intelligence, the auto-generation of structured, graphical process models based on unstructured input data has re-surged as a line of research ~\cite{ConversationalProcessModelling:Stateof,ProcessModelGenerationfromNatural,GeneratingBPMNdiagramfromtextual,ExtractingBusinessProcessEntitiesand,Leveragingpretrainedlanguagemodelsfor,LargeLanguageModelsCanAccomplish}. Here, the main input modality is text. However, unstructured process models are in practice often persisted in image-based or image-like formats (e.g., PNG or PDF files) containing flow charts as well as corresponding textual information (e.g., a process description).
Hence, extending the input modality (using multimodal generative models) to handle combined image-based and textual input yields promising possibilities.
In contrast to existing approaches, where the image modality is handled separately using a traditional Convolutional Neural Network (CNN) (cf.~\cite{IEEEXploreFullTextPDF,AutomaticRecognitionofBusinessProcess,paper_246,TowardsCreatingBusinessProcessModelsb}), multimodal approaches offer several advantages: First, only one model is required, reducing both maintenance and development efforts. Second, generative AI approaches, Large Language Models (LLMs) in particular, are known for their transfer learning capabilities. The latter allow handling different tasks, for example, various types of diagrams, better compared to traditional CNNs trained for one specific purpose~\cite{LanguageModelsareFewShotLearners,MME:AComprehensiveEvaluationBenchmark}.

Therefore, this paper presents an in-depth investigation of multimodal generative models to auto-generate graphical process models from text and (graphical) figures. We focus on providing a systematic evaluation approach to ensure a scientific comparison and consider future model developments and improvements. We further introduce a small dataset and a set of metrics for ground truth-based evaluation. Based on this dataset, we evaluate commercial GPT capabilities using zero-, one-, and few-shot prompting strategies and discuss their implications.


\section{Background \& Related Work}
\label{sec:preliminaries}
Applying LLMs in the BPM domain is considered a promising research direction. Vidgof et al. outline the opportunities and challenges of LLMs for the BPM domain~\cite{LargeLanguageModelsforBusiness}. Among others, they suggest using multimodal LLMs to produce process models automatically from documentation. Similarly, Kampik et al. speculate that LLMs can significantly reduce effort and lower knowledge barriers in the BPM domain~\cite{LargeProcessModels:BusinessProcess}. Beyond LLMs, Van der Aa et al. outline the opportunities and challenges of applying NLP in BPM~\cite{ChallengesandOpportunitiesofApplying} and Dumas et al. provide a vision of AI-augmented Business Process Management Systems and possible research questions to achieve it~\cite{AIaugmentedBusinessProcessManagementSystems}. Beheshti et al. propose opportunities for training an LLM on business process data to achieve various BPM tasks~\cite{ProcessGPT:TransformingBusinessProcessManagement}. Similarly, Rizk et al. suggest a process-specific Foundational Model and the opportunities it could offer~\cite{ACaseforBusinessProcessSpecific}. These position papers offer first intuitions and recaps of the state of the art, but feasibility assessments remain vague.

A nascent body of research exists regarding the generation process models from text with NLP in general and GPTs/LLMs in particular. Klievtsova et al. evaluated application scenarios for modeling business processes with LLMs based on textual conversations on a real-world data set~\cite{ConversationalProcessModelling:Stateof}. Bilal et al. found 11 NLP and 8 BPMN tools for generating models from textual requirements utilizing NLP~\cite{AComprehensiveInvestigationofBPMN}. Friedrich et al. and Sholiq et al. evaluated NLP techniques to generate process models from natural language~\cite{ProcessModelGenerationfromNatural,GeneratingBPMNdiagramfromtextual}. Bellan et al. use LLMs for extracting business process entities and relations from texts~\cite{ExtractingBusinessProcessEntitiesand,Leveragingpretrainedlanguagemodelsfor}. Grohs et al. apply LLMs to the generation of both imperative and declarative process models from textual descriptions~\cite{LargeLanguageModelsCanAccomplish}. 

Regarding the use of images to generate process models, Schaefer et al. introduce a CNN model to create BPMN models from hand-drawn sketches~\cite{IEEEXploreFullTextPDF}.  Kang et al. transform BPMN business process images into Petri nets with traditional deep learning techniques~\cite{AutomaticRecognitionofBusinessProcess}. Antinori et al. recreate business process models from BPMN images using traditional object recognition and Optical Character Recognition (OCR) methods~\cite{paper_246}. Gantayat et al. create BPMN models from images using traditional CNNs~\cite{TowardsCreatingBusinessProcessModelsb}.

In contrast, our approach explores a multimodal context where both images and text converge within documents. Additionally, we leverage multimodal LLMs, diverging from the methods highlighted in the preceding papers.
This approach offers greater versatility since an LLM can perform multiple tasks using fundamental reasoning skills without being exclusively trained on a specific type of image or text.

To the best of our knowledge, no previous work has investigated the capabilities of multimodal generative models to auto-generate graphical process models from intersected text- and image-based inputs in detail. The paper aims to contribute towards closing this gap.






\section{Method}
\label{sec:method}
We initially present a small dataset and evaluation metrics for ground truth-based assessment of multimodal process model generation. Subsequently, we employ commercial GPT capabilities for the generation process. The complete source code, including the dataset, evaluation metrics, and generation part, is publicly available \footnote{\url{https://github.com/SAP-samples/multimodal-generative-ai-for-bpm}}.

\subsection{Dataset}
\label{subsec:dataset}
\subsubsection{Creation}
Although several datasets of process models ~\cite{SAPSignavioAcademicModels:A,camundabpmnforresearch:Acollectionof}, model-text pairs ~\cite{MaD:ADatasetforInterviewbased,PET:AnAnnotatedDatasetfor,ConversationalProcessModelling:Stateof,Nataliiaconversational_modelingGitLab} and model-image pairs ~\cite{IEEEXploreFullTextPDF,paper_246} exist, none contains multimodal process documentation and respective ground truths. Therefore, we augment a subset of the SAP-SAM dataset, which contains around a million of process models created by researchers, teachers, and students~\cite{SAPSignavioAcademicModels:A}. The SAP-SAM dataset is filtered for high-quality and representative models. These process models are then parsed into a simplified JSON and, consequently, serve as ground truth process models. Afterwards, the cleaned models are used to generate multimodal process documentation in a PDF format with the
SAP Signavio Process Manager, a commercial process management tool\footnote{Cf. \url{https://www.signavio.com/products/process-manager/}; accessed at 23-03-2023.}.
Resulting PDFs are then converted to corresponding PNG replicas.

\subsubsection{Characteristics}
Overall, the dataset comprises 123 models, including original SAP-SAM data, multimodal process documentation in both PDF and PNG formats as well as the ground truth in JSON format. In a nutshell, the documentation comprises a cover page, a table of contents, the meta-information, a BPMN diagram, and short descriptions for each element.  Depending on the process model, the documentation spans between 4-18 pages. Typically, the textual descriptions match the visual models, while occasionally additional information or a brief overview is provided.

The JSON ground truth schema consists of different objects, each referenced by corresponding IDs. Objects may represent a task, an event, a gateway, a pool, a lane, a message flow, or a sequence flow. Tasks, events, and gateways have a name and a type, while the name is optional for gateways. A pool has a name and can have multiple lanes. A lane has a name and can contain multiple referenceable objects. A message flow has an optional label and exactly one source and one target referencing objects. Sequence flows are defined equivalent to the message flows, although the label attribute is called a condition. Overall, the schema covers the most used BPMN elements. It includes all elements used in more than 25\% of publicly available repositories examined by Compagnucci et al.~\cite[pp.~87--88]{TrendsontheUsageof}. Covering all elements used in more than 3\% of models, the schema can be extended by adding new lists for associations and object elements with the respective types data object, text annotation, message, group, and a flow type for default flow.

The number of ground truth elements in the dataset varies between 13 and 88, with an average of 39. This illustrates the adherence of the model to the guideline of limiting elements to 50 per model, suggesting high quality~\cite{MENDLING2010127}. Predominantly, Tasks and Sequence Flows constitute the elements, featuring 10 task types with abstract and manual tasks being most common. Moreover, the dataset includes 17 event types, notably the End None and Start None Events, and identifies four gateway types, with exclusive and parallel gateways being favored. These findings closely match the distributions reported by Compagnucci et al.~\cite[pp.~87--88]{TrendsontheUsageof}, demonstrating significant real-world representativeness.

\subsection{Evaluation Framework}
\label{subsec:framework}

\subsubsection{Element based breakdown}
First, the ground truth and to-be-evaluated models (generated models) are broken down into individual elements. We define the following multisets: $TN$ as task names, $TT$ as task types, $EN$ as event names, $ET$ as event types, $GN$ as gateway names and $GT$ as gateway types. Further, we define $LN$ as a multiset of lane names. A lane name is defined as a tuple $(p, l)$ where $p$ is the pool label and $l$ is the label of the lane. And $LR$ is a multiset of lanes with reference. A lane with reference is a triple $(p, l, r)$ where $p$ is the pool label, $l$ is the label of the lane, and r $\in$ $ TN \cup EN \cup GT$. $GT$ is chosen over $GN$ because the names of the gateways are optional, while the type is mandatory. In addition, let $SF$ be a multiset of sequence flows. A sequence flow is defined as a triple $(s, c, t)$ where $s, t$ $\in$ $TN \cup EN \cup GT \cup LN$ and $c$ provides a name for the sequence flow. Equivalent, let $MF$ be a multiset of message flows. A message flow is defined as a triple $(s, l, t)$ where $s, t$ $\in$ $ TN \cup EN \cup GT \cup LN$ and $l$ provides a name for the message flow. The model is then described as a joint set of elements:
\begin{align}
    M = \{TN, TT, EN, ET, GN, GT, LN, LR, SF, MF\}
\end{align}

As a short-hand, the elements are referred to as $E_i$:
\begin{align}
M = \{E_i : 1 \leq i \leq 10\}
\end{align}

\subsubsection{Score calculation}
Next, the similarity scores of the two models are calculated. Let $M_{1} = \{E_{i,1} : 1 \leq i \leq 10\}$ be the ground truth model and $M_{2}= \{E_{i,2} : 1 \leq i \leq 10\}$ be the generated model. The overall similarity score is calculated as follows:

\begin{align}
    \text{sim}(M_{1}, M_{2}) &= \frac{\sum_{i=1}^{10} w_i 
    \cdot dice_{SFA}(E_{i,1}, E_{i,2})}{\sum_{i=1}^{10} w_i} \quad \text{with } w_i = |E_{i,1}| + |E_{i,2}|
\end{align}

where $dice_{SFA} (E_{i,1}, E_{i,2})$ is an element-wise adjusted Sørensen–Dice coefficient and the weights are element cardinalities. The original Sørensen–Dice coefficient is defined as~\cite{295_SrensenThorvald,MeasuresoftheAmountof}:
\begin{equation}
dice(E_{i,1},E_{i,2}) = \frac {2 |E_{i,1} \cap E_{i,2}|}{|E_{i,1}| + |E_{i,2}|}
\end{equation}

Further aggregated scores for tasks ($TN$ and $TT$), events ($EN$ and $ET$), gateways ($GN$ and $GT$), flows ($SF$ and $MF$) and lanes ($LN$ and $LR$) are calculated analogously to the overall score.

\subsubsection{$dice_{SFA}$} $dice_{SFA}$ is a semantic- and frequency-aware adjustment of the Sørensen–Dice coefficient to calculate the similarity of two lists. First, semantically similar items from the two lists are matched, even when syntactically dissimilar. To achieve that, a BERT sentence transformer for item vectorization is employed and items are considered similar if their cosine similarity exceeds a 0.7 threshold. Note that this threshold has been determined experimentally, and may be  adjusted for different contexts. Items are only matched once, prioritizing matches with the highest similarity first. Next, indexing items enables the Sørensen–Dice coefficient to process multisets by converting them into sets with preserved entries. Finally, the original Sørensen–Dice coefficient is calculated. Note that the corresponding source code of the evaluation framework is provided in GitHub \footnote{\url{https://github.com/SAP-samples/multimodal-generative-ai-for-bpm}}.

\subsection{Generation Process}
\label{subsec:generation}
We leverage a multimodal LLM to generate process models from the process documentation in the dataset. We decided on GPT-4V because it performed best in benchmarks of the following categories: existence recognition, element counting, position understanding, OCR, common sense reasoning, and code reasoning~\cite{MME:AComprehensiveEvaluationBenchmark}.

Listing \ref{lst:metaprompt} describes the meta prompt used for the LLM and follows different prompting strategies. First, it applies priming: ``You are a BPMN expert.'' Then, it clearly and precisely describes the to-be-performed task: extracting the process information from the passed document. Next, the prompt constrains the output to only generate the JSON according to the passed schema. The schema corresponds to the defined ground truth schema. The meta prompt has been refined based on iterative experiments. For example, explicitly mentioning to add the sequence and message flow resulted from them often being empty. We generated the models in three different scenarios~\cite{LanguageModelsareFewShotLearners}:
\begin{enumerate}
    \item Zero-Shot Prompting:  Giving the LLM only the meta prompt and no example at all to perform the task.
    \item One-Shot Prompting: Providing the LLM the meta prompt and one example to perform the task.
    \item Few-Shot Prompting: Providing the LLM the meta prompt and three examples to perform the task.
\end{enumerate}

The first three models of the dataset are used as examples of the one- and few-shot
approach. Each approach is then applied to the remaining 120 models.

\begin{lstlisting}[language=Python, caption=Meta prompt, captionpos=b, breaklines]
meta_prompt = """### Instruction ###
  You are a BPMN expert. Your task is to extract process information out of
  the passed documents which are parsed as a list of images where each image
  represents one page of the document. Make sure that you include the
  sequence and message flow. Use numbers for the ids starting from zero.
  Generate json according to the following schema for extracting the process
  information. Only output the generated json.
  
  ### Schema ###
  """ + str(bpmn_schema)
\end{lstlisting}
\label{lst:metaprompt}


\section{Results} 
\label{sec:results}
\begin{figure}[h!]
	\centering
	\includegraphics[width=0.9\textwidth]{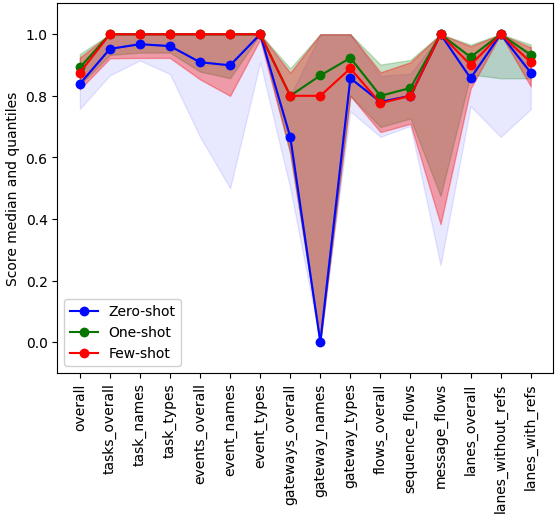}
	\caption{Median scores and quantiles for zero-, one-, and few-shots.} \label{fig:score_comparison}
\end{figure}

Figure \ref{fig:score_comparison} visualizes the score distributions with medians as the central tendency along with shaded areas between 25\% and 75\% quantiles which represent score variability. We have chosen to show the medians and quantiles because the score distributions are skewed. Additionally, Table \ref{table:score_comparison} lists the score averages. The overall score median for the one-shot scenario is 89\%, for the few-shot scenario 87\%, and 83\% for the zero-shot scenario. For all approaches, the score for task names, task types, and event types is higher than the overall score, while the event names are slightly below it with larger variability. The largest variability is seen in gateway names which also has the lowest similarity score. Message flow has higher score variability than sequence flow. The overall score for flows lies slightly below the overall score for all elements.  

The performance of the one-shot and few-shot scenario is very similar, and both significantly outperform the zero-shot scenario and reduce variability in scores. On average, the difference between one-shot and few-shot is below 1 percent point, while the difference to few-shot is around 7 - 8 percent points.

\begin{table}[!h]
\centering
\begin{tabulary}{\linewidth}{|L|C|C|C|}
\hline
\textbf{Score Name} & \textbf{Zero-Shot} & \textbf{One-Shot} & \textbf{Few-Shot} \\ \hline
Overall & 0.812654 & 0.871079 & 0.861744 \\ \hline
Tasks Overall & 0.889146 & 0.939833 & 0.944912 \\ \hline
Task Names & 0.919780 & 0.958201 & 0.946046 \\ \hline
Task Types & 0.858431 & 0.922100 & 0.944000 \\ \hline
Events Overall & 0.814431 & 0.908772 & 0.893474 \\ \hline
Event Names & 0.697021 & 0.865807 & 0.848409 \\ \hline
Event Types & 0.924640 & 0.945658 & 0.932955 \\ \hline
Gateways Overall & 0.609449 & 0.696482 & 0.686032 \\ \hline
Gateway Names & 0.284444 & 0.575058 & 0.601764 \\ \hline
Gateway Types & 0.856959 & 0.883706 & 0.865670 \\ \hline
Flows Overall & 0.736907 & 0.776942 & 0.751045 \\ \hline
Sequence Flows & 0.750651 & 0.792892 & 0.770219 \\ \hline
Message Flows & 0.682600 & 0.723168 & 0.696521 \\ \hline
Lanes Overall & 0.813442 & 0.881780 & 0.868453 \\ \hline
Lane Names & 0.825298 & 0.872698 & 0.878962 \\ \hline
Lane Refs & 0.807339 & 0.877842 & 0.859720 \\ \hline
\end{tabulary}

\setlength{\abovecaptionskip}{10pt}
\caption{Comparison of average zero-, one-, and few-shot scores.}
\label{table:score_comparison}
\end{table}


\section{Discussion}
In this section, we discuss the implications and limitations of our study and its results.
\subsection{Implications}
The results, with an average overall similarity score of 87 percent for the one-shot approach, indicate that creating process models from multimodal documents using generative AI techniques is feasible. However, human human feedback might still be required to improve the results.
Our experiments provide first evidence that a multimodal GPT approach can be applied, as an alternative to separately processing text and images, for process model generation.
In the multimodal case, only one machine learning model is required that may then also be used for alternative tasks, thus reducing both development and maintenance efforts. 
In addition, the multimodal GPT does not require effortful special-purpose training but can be used ``off-the-shelf''.
We expect that future enhancements, including the introduction of function-calling capabilities and advancements in models like Gemini 1.5, along with optimized prompting, can further solidify evidence regarding the feasibility of multimodal GPTs for automated process model generation based on unstructured documents. 

For the task at hand, GPTs can effectively handle JSON schemas and one-shot prompting often suffices without additional examples to significantly improving outcomes compared to zero-shot prompting. While GPTs excel at identifying tasks and events, they struggle with flows, possibly due to relational complexity.
We further recognize challenges related to identifying gateway labels. A possible reason might be that the presence of gateway labels is optional and label placement varies substantially.

\subsection{Limitations}

Several factors limit the results of the presented study. First, the relative contribution of the diagram images and textual information to the overall score has not been analyzed.
Second, the study exclusively utilized one LLM model, limiting insights into potential variations in performance or challenges with other models. Rapid advancements in technology, evidenced by the release of GPT-4V and Gemini 1.5 during the paper period, highlight the dynamic nature of the field and potential obsolescence. A stable benchmarking environment was established to address this, facilitating future technological comparisons.
Third, resource constraints prevented fine-tuning strategies, and prompt optimization could further be enhanced with additional effort.
Finally, while empirically somewhat robust, the evaluation methodology is imperfect. Two specific issues arise: First, employing BERT for similarity assessments with a fixed threshold may lead to potential deviations from linearity. Second, set overlap results in unequal weighting of elements, presenting an additional challenge due to the ambiguity in determining the appropriate priority or weight for each element. Despite these intricacies, the empirical data consistently affirm the methodology's intuitive precision and evaluation uniformity. However, future work can potentially further enhance the evaluation approach, for example by considering existing research on process model similarity assessment~\cite{DIJKMAN2011498}.

\section{Conclusion}
\label{sec:conclusion}
The paper presented an investigation of the capabilities of
Generative Pre-trained Transformers (GPTs) to auto-generate graphical
process models from multi-modal (i.e., text- and image-based) inputs.
We first created a dataset of 123 models containing both process documentations and ground truths. Then, we introduced a ground truth-based evaluation framework incorporating an element-wise breakdown, semantic and frequency awareness as well as the Sørensen–Dice coefficient. We evaluated GPT-4V capabilities using zero-, one-, and few-shot prompting strategies. Our results, with an average evaluation score of 87\%, indicated that GPTs can be useful tools for semi-automated process modeling based on multi-modal
inputs. In addition, the dataset, the evaluation metrics, and the open-source evaluation code provide a structured framework for continued systematic evaluations moving forward.

Future research should benchmark text-only and image-only models against the dataset to evaluate their performance relative to traditional methods. Additionally, analyzing  the relative contribution of text and image information to the generated process model could be valuable. Extending the ground truth to include a wider array of BPMN elements, could further improve generated models. Expanding the dataset to encompass a broader variety of diagrams, images, and textual descriptions could also enrich the benchmarking possibilities. Moreover, new models and approaches can be benchmarked using the introduced dataset. Exploring chain-of-thought prompting, function calling, and fine-tuning also seems promising, though resource-intensive.

\bibliographystyle{splncs04}
\bibliography{main}

\FloatBarrier

\end{document}